\documentclass[aps,prl,reprint,superscriptaddress]{revtex4-1}
\usepackage {graphicx}
\usepackage{bm}      
\usepackage{amssymb}   
\usepackage[caption=false]{subfig}

\begin{document}

\title{
Electron-Electron Interactions and the Paired-to-Nematic 
Quantum Phase Transition in the Second Landau Level}

\author{K.A. Schreiber}
\affiliation{Department of Physics and Astronomy, Purdue University, West Lafayette, Indiana 47907, USA}
\author{N. Samkharadze}
\email[]{Present address: QuTech and Kavli Institute of NanoScience, Delft University of Technology,  
Lorentzweg 1, 2628 CJ Delft, Netherlands.}
\affiliation{Department of Physics and Astronomy, Purdue University, West Lafayette, Indiana 47907, USA}
\author{G.C. Gardner}
\affiliation{School of Materials Engineering, Purdue University, West Lafayette, Indiana 47907, USA}
\affiliation{Birck Nanotechnology Center Purdue University, West Lafayette, Indiana 47907, USA}
\author{Y. Lyanda-Geller}
\affiliation{Department of Physics and Astronomy, Purdue University, West Lafayette, Indiana 47907, USA}
\author{M.J. Manfra}
\affiliation{Department of Physics and Astronomy, Purdue University, West Lafayette, Indiana 47907, USA}
\affiliation{School of Materials Engineering, Purdue University, West Lafayette, Indiana 47907, USA}
\affiliation{Birck Nanotechnology Center Purdue University, West Lafayette, Indiana 47907, USA}
\affiliation{School of Electrical and Computer Engineering, Purdue University, West Lafayette, Indiana 47907, USA }
\author{L.N. Pfeiffer}
\affiliation{Department of Electrical Engineering, Princeton University, Princeton, New Jersey 08544, USA}
\author{K.W. West}
\affiliation{Department of Electrical Engineering, Princeton University, Princeton, New Jersey 08544, USA}
\author{G.A. Cs\'{a}thy}
\email[]{Corresponding author: gcsathy@purdue.edu}
\affiliation{Department of Physics and Astronomy, Purdue University, West Lafayette, Indiana 47907, USA}
\affiliation{Birck Nanotechnology Center Purdue University, West Lafayette, Indiana 47907, USA}

\date{\today}

\begin{abstract}

In spite of its ubiquity in strongly correlated systems, the competition of paired and nematic
ground states remains poorly understood. Recently such a competition was reported in the 
two-dimensional electron gas at filling factor $\nu=5/2$. At this filling factor
a pressure-induced quantum phase transition was observed from the paired 
fractional quantum Hall state to the quantum Hall nematic.
Here we show that the pressure induced paired-to-nematic transition 
also develops at $\nu=7/2$, demonstrating therefore
this transition in both spin branches of the second orbital Landau level. 
However, we find that pressure is not the only parameter controlling this transition.
Indeed, ground states consistent with those observed under pressure also develop in a sample measured
at ambient pressure, but in which the electron-electron interaction was tuned close to its 
value at the quantum critical point. Our experiments 
suggest that electron-electron interactions play a critical role in driving
the paired-to-nematic transition.

\end{abstract}

\maketitle

{\bf Introduction.} 

Nematicity is of interest in various strongly correlated electron systems 
\cite{kivelson,lilly,du,fogler,moessner,kiv,fradkin}. It is generally accepted that nematicity originates from 
competing interactions on different length scales. However,  the interplay of nematicity with 
other phases, such as with superconductivity in the cuprates \cite{high1,high2}, is not understood. 
For example, the influence of the nematic fluctuations on  
pairing in the superconductive phase is actively debated \cite{fluct1,fluct2,fluct3,fluct4,kim17}.

Nematic and paired ground states also develop in half-filled Landau levels of the
two-dimensional electron gas confined to high quality GaAs/AlGaAs structures.
Indeed, a strong resistance anisotropy at the Landau level filling factors $\nu=9/2, 11/2, 13/2, ...$
signals a ground state state with broken rotational symmetry \cite{lilly,du}.
There are two distinct ground states consistent with such an anisotropy:
the smectic and nematic phases \cite{kiv,fradkin,fogler,moessner,nem7}. The difference between these two is that the
former has unidirectional translational order, whereas the latter does not. Since electrons are buried deep within the
GaAs crystal, a direct detection of translational order remains elusive. In addition,
disorder is expected to destroy translational order and therefore it favors nematicity.
In lack of certainty about the translational order, these anisotropic phases
are often referred to as the quantum Hall nematic, or simply the nematic. Henceforth we adopt this terminology.

In contrast to the ground states at filling factors $\nu=9/2, 11/2, 13/2,...$, those at $\nu=5/2$ and $7/2$
are isotropic fractional quantum Hall states (FQHSs) \cite{firstfivehalf,fivehalf,sevenhalf,MooreRead}.
Since in GaAs each orbital Landau level has two spin branches, $\nu=5/2$ and $7/2$
describe two half-filled spin banches of the second orbital Landau level. 
FQHSs are incompressible and possess topological order. 
Topological order in the FQHSs at $\nu=5/2$ and $7/2$ is still under active investigation. 
However, within the framework of the composite fermion theory \cite{jain,halperin}, these FQHSs 
are due to pairing of the composite fermions, hence the paired FQHS terminology \cite{pair1,pair2,pair3,pair4}. 
A schematic representation of the ordered ground states at half-filling is seen in Fig.1.

\begin{figure}[!b]
\centering
\includegraphics[width=0.95\columnwidth]{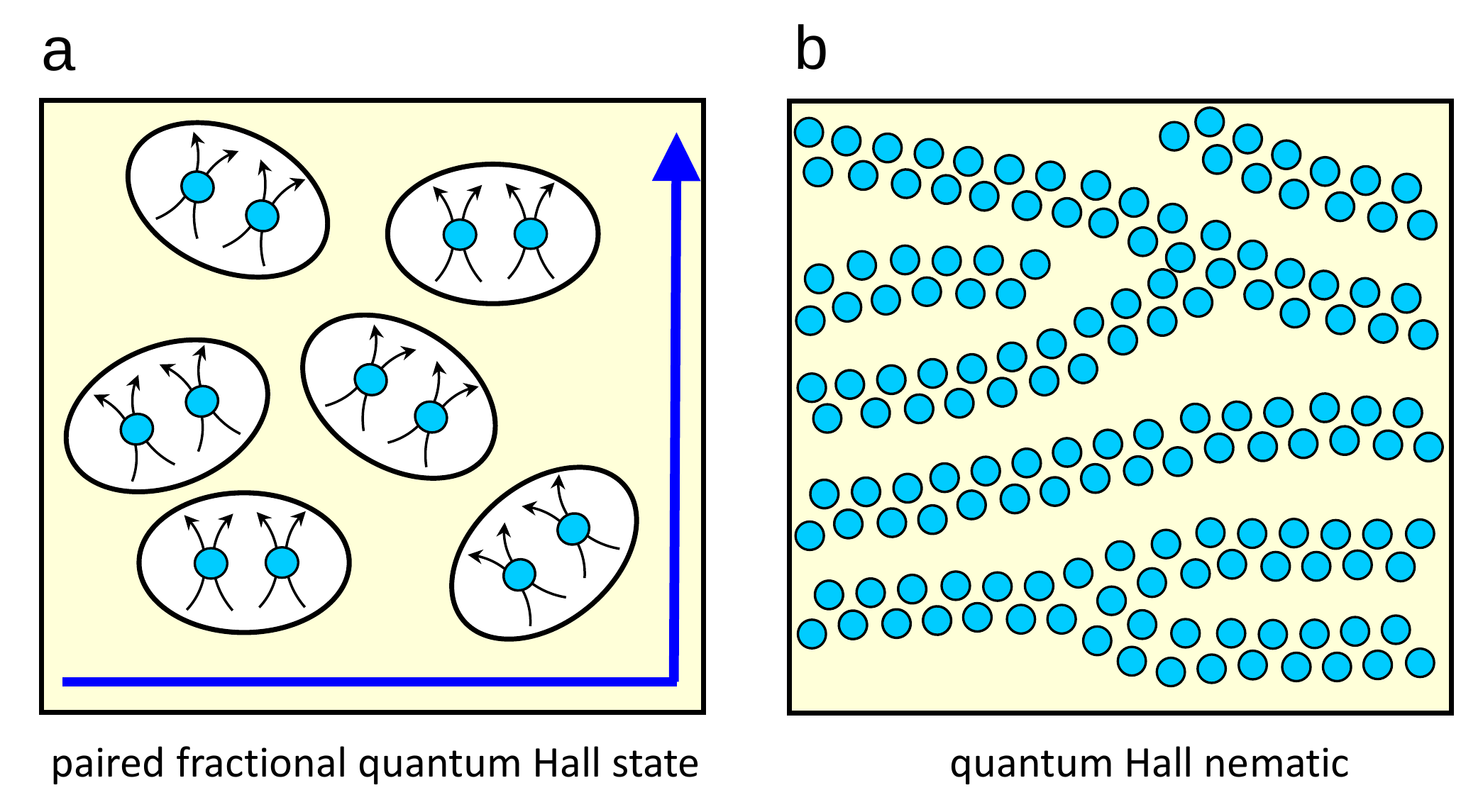}
\centering
\caption{{\bf Schematic of the ordered phases at even denominator
filling factors.} {\bf a} The paired FQHS 
consists of Cooper pairs of composite fermions and posesses edge states \cite{pair1,pair2,pair3,pair4}. 
Composite fermions are depicted as electrons with two magnetic fluxlines attached \cite{jain}. 
{\bf b} The quantum Hall nematic is a filamentary electronic phase which
breaks rotational symmetry \cite{kiv}. 
}
\label{fig-1}
\end{figure}

\begin{figure*}
\centering
\includegraphics[width=0.95\textwidth]{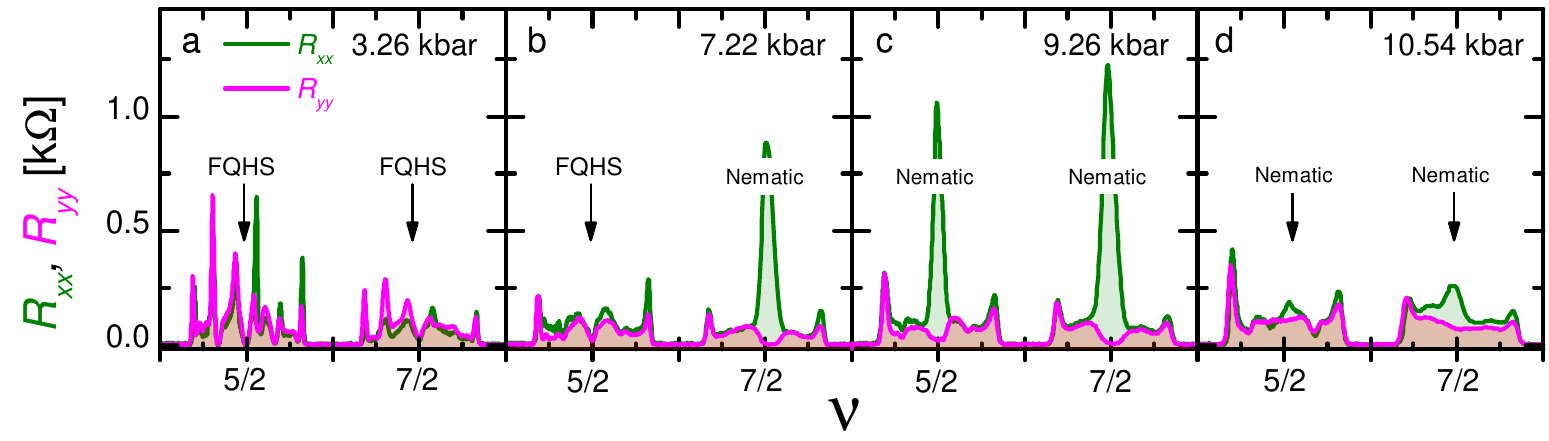}
\centering
\caption{ {\bf Pressure dependence of magnetoresistance in the second Landau level.} 
$R_{xx}$ and $R_{yy}$ are shown for Sample A,
as measured along two mutually perpendicular crystal axes of GaAs.
Pressures are 3.26, 7.22, 9.26, and 10.54~kbar for panels {\bf a}, {\bf b}, {\bf c}, and {\bf d}, respectively.
The temperature is $T \approx 12$ mK. Traces allow us to observe the evolution of ground states at
half-filled Landau levels $\nu=5/2$ and $\nu=7/2$. The nature of the ground state
at these two filling factors is marked by text; FQHS stands for fractional quantum Hall state.
}
\label{fig-2}
\end{figure*}

At a given orbital quantum number, the ordered ground state at half-filled Landau level
is either the nematic or the FQHS, but a transition between them
did not seem possible in the absence of a symmetry breaking field favoring the nematic.
This was surprising, since in the second orbital Landau level at $\nu=5/2$ and $7/2$
tilted field experiments suggested that the two ground states are close in energy \cite{tilt1,tilt2}. 
Additionally, incipient nematicity was seen at $\nu=7/2$ \cite{Pan72}. However,
a phase transition from the FQHS to the nematic in the absence of an 
in-plane symmetry breaking magnetic field was only recently observed \cite{Samkharadze}. 
In these experiments the transition occurred at $\nu=5/2$ and it was driven by pressure. 
Because of the hydrostatic nature
of the applied pressure, the rotational symmetry in these experiments was not explicitly broken. 

Our understanding of the paired-to-nematic phase transition 
and the associated quantum critical point remains lacunar. Tuning the Haldane
pseudopotentials in the second Landau level induces a transition from the paired FQHS to the nematic \cite{pair4}. 
However, the very nature of this transition remains unknown. Recent theories find that the
nematic phase is stabilized by a Pomeranchuk instability of the Fermi sea of composite fermions 
\cite{frad,rez18}. A paired-to-nematic transition is compatible with these theories,
but details have not yet been worked out. In another work, the influence of the nematic
fluctuations on the paired FQHS has been examined, with the assumption that a paired-to-nematic 
transition exists \cite{kim17}.  Ref.\cite{inti17} captures a paired-to-nematic transition 
by tuning the mass anisotropy of the carriers. 
However, there is no evidence that such a mass anisotropy plays a significant role in the electron gas
hosted in GaAs. What determines the quantum critical point? 
Can the transition be induced using a parameter other than pressure?

Guided by these questions, here we investigate the ground state of the two-dimensional electron gas 
in a wider phase space. We establish that the paired-to-nematic transition also occurs 
at filling factor $\nu=7/2$, the particle-hole conjugate of $\nu=5/2$. However, this transition is
not observed outside the second Landau level nor at unpaired FQHSs forming in the second Landau level.
This finding highlights the importance of pairing in the transition from a FQHS to the nematic
and establishes the presence of the paired-to-nematic transition and the associated quantum critical point
in both spin branches of the half-filled second orbital Landau level.
We observe that the critical pressure of the transition at $\nu=7/2$ is much reduced when compared to that
at $\nu=5/2$. In contrast, we find that the transition occurs at nearly the same magnetic field.
This observation allows us to conclude that pressure is not a primary driver of the transition, but the
electron-electron interaction is. To demonstrate this,
we show that ground states consistent with those at high pressures also develop in
at {\it ambient pressure}, but in which the electron-electron interaction is engineered to be
close to its critical value. 

{\bf Results.}

{\bf Samples.} We measured two samples.
Sample A is a 30~nm quantum well sample with an as-grown density of $29.0 \times 10^{10}$ cm$^{-2}$, which
was investigated under hydrostatic pressure. The mobility of this sample in the 
ambient is $20 \times10^{6}$ cm$^2$V$^{-1}$s$^{-1}$. Sample B  is also a 30~nm quantum well sample,
but with an as-grown density of $10.9 \times 10^{10}$ cm$^{-2}$. Sample B
was measured only at ambient pressure and has a mobility of $18 \times10^{6}$ cm$^2$V$^{-1}$s$^{-1}$.

\begin{figure*}[!htb]
\centering
\includegraphics[width=0.9\textwidth]{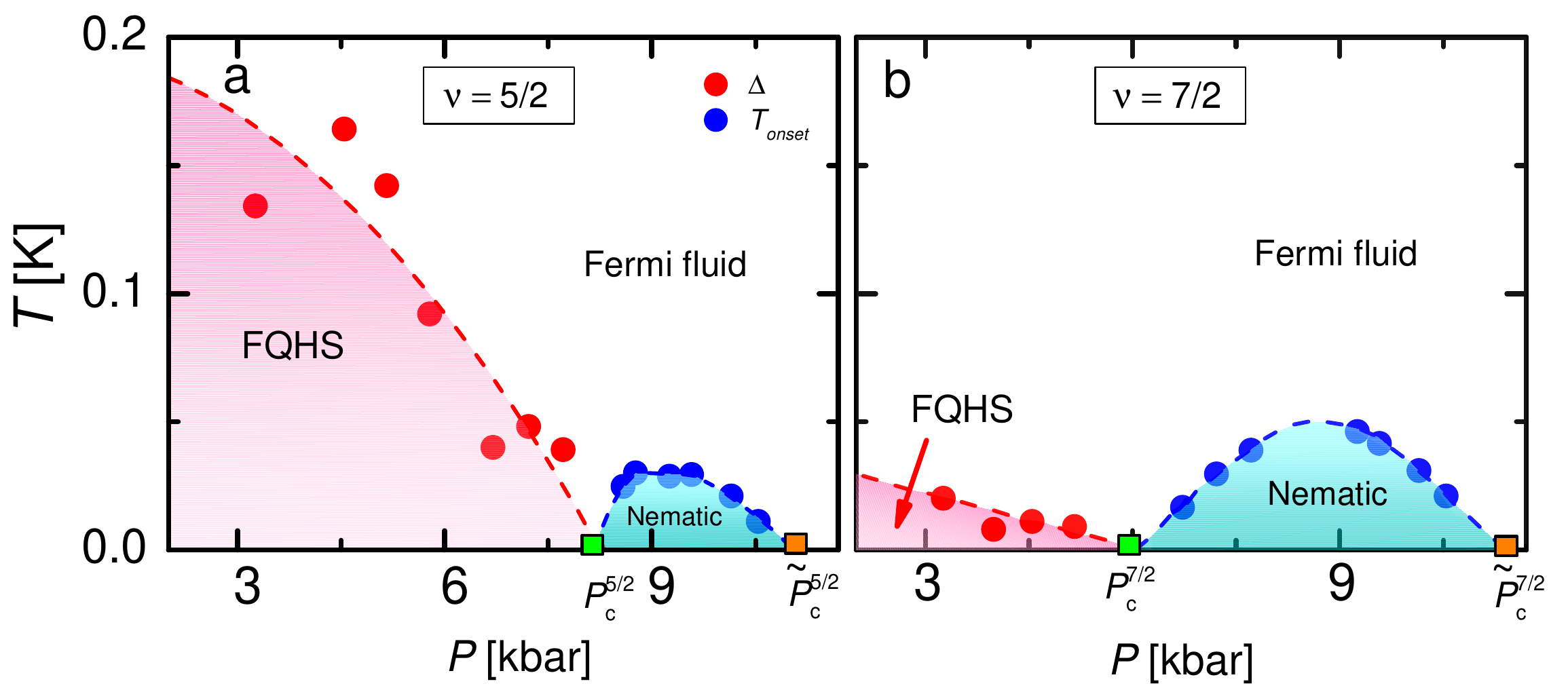}
\caption{{\bf Stability diagrams in the half-filled second Landau level.}
The stability diagrams for Sample A are obtained from plotting the
energy gap $\Delta$ of the FQHS and the onset temperature $T_{onset}$
of the nematic. Analysis is done at $\nu=5/2$ (panel {\bf a} )and $\nu=7/2$ (panel {\bf b}).
The energy gaps decreases with increasing pressure,
while the onset temperature of the nematic exhibits a dome-like shape.
Green squares represent the quantum critical point of the paired-to-nematic and
orange squares of the nematic-to-Fermi fluid transition. Lines are guides to the eye.
}
\label{fig-3}
\end{figure*}

{\bf Terminology.}
The energy spectrum of a two-dimensional electron gas of density $n$ 
in a magnetic field $B$ at large enough fields consists of spin-split
Landau levels. The number of filled energy levels is given by
the Landau level filling factor $\nu=h n / e B$, where $e$ is the electron charge and $h$ is Planck's constant. 
In the absence of the valley degree of freedom,
the second orbital Landau level in GaAs corresponds to the $2< \nu < 4$ range.
Of this range, the $2< \nu < 3$ is the lower spin branch, while
the $3< \nu < 4$ range the upper spin branch. Therefore at $\nu=5/2$ and $\nu=7/2$ the system has
half-filled Landau levels with the same orbital quantum number, but different spin quantum numbers.

{\bf Pressure dependent magnetoresistance at low temperatures.} 
Figure 2 highlights the evolution of the magnetoresistance in the two spin branches of the second orbital Landau level
at the lowest temperature of  $T \approx 12$~mK reached in our pressure cell.
Traces are measured along two mutually perpendicular directions: 
$R_{xx}$ along the $[1\bar{1}0]$ and $R_{yy}$ along the $[110]$ crystal axis of GaAs.
These traces show several features which can be associated with known 
ground states of the electron gas at ambient pressure \cite{Xia,Ethan};
in the following we focus our attention to $\nu=5/2$ and $\nu=7/2$.
The magnetoresistance at $\nu=5/2$ is isotropic and vanishing at $3.26$
and $7.22$~kbar, signaling a FQHS \cite{firstfivehalf,fivehalf}. The magnetoresistance at $\nu=5/2$ is
strongly anisotropic at $9.26$~kbar and has very little anisotropy at $10.54$~kbar,
exhibiting therefore nematic behavior \cite{lilly,du}.
This behavior with increasing pressure is consistent with 
a FQHS, quantum Hall nematic, isotropic Fermi fluid sequence of ground states \cite{Samkharadze}.

The magnetoresistance trend at $\nu=7/2$ shown in Fig.2
is qualitatively similar to that at $\nu=5/2$ as it evolves  
from isotropic and nearly vanishing at $3.26$~kbar, to strongly anisotropic
at $7.22$ and $9.26$~kbar, to weakly anisotropic at $10.54$~kbar. 
This behavior at $\nu=7/2$ suggests the same sequence of ground states as at $\nu=5/2$
and hints at the existence of a paired-to-nematic transition at $\nu=7/2$.  
A FQHS at at $\nu=7/2$ and at $3.26$~kbar is supported by observation of 
Hall resistance quantization, shown in Supplementary Figure 1.
Furthermore, as demonstrated by temperature-dependent measurements shown in
Supplementary Figure 2, the nematic observed at $\nu=7/2$ is a compressible ground state
similar to the one observed at $\nu=9/2$ of the third Landau level in samples measured
at ambient pressure \cite{lilly,du}.

At certain pressures, Fig.2 shows the same type of ground states
at both $\nu=5/2$ and $7/2$. Indeed, at $P=3.26$~kbar we observe two FQHSs,
while at $P=9.26$ and $10.54$~kbar we observe two nematic phases. 
This arrangement of similar ground states at different half-filled spin branches of
a given orbital Landau level is typical for samples in the ambient.
For example, ground states at both $\nu=5/2$ and $7/2$ in the second Landau level
are FQHSs \cite{sevenhalf} and those at $\nu=9/2$ and $11/2$ in the
third Landau level are nematic states \cite{lilly,du}. 
At $P=7.22$~kbar, however, we observe an exception to such an arrangement. Indeed, at this pressure
the ground state at $\nu=5/2$ is a FQHS, while that at $\nu=7/2$ is the nematic.
This asymmetry implies that the nematic at $\nu=7/2$ is stabilized at a lower pressure than that at $\nu=5/2$.

\begin{figure*}[!htb]
\centering
\includegraphics[width=0.9\textwidth]{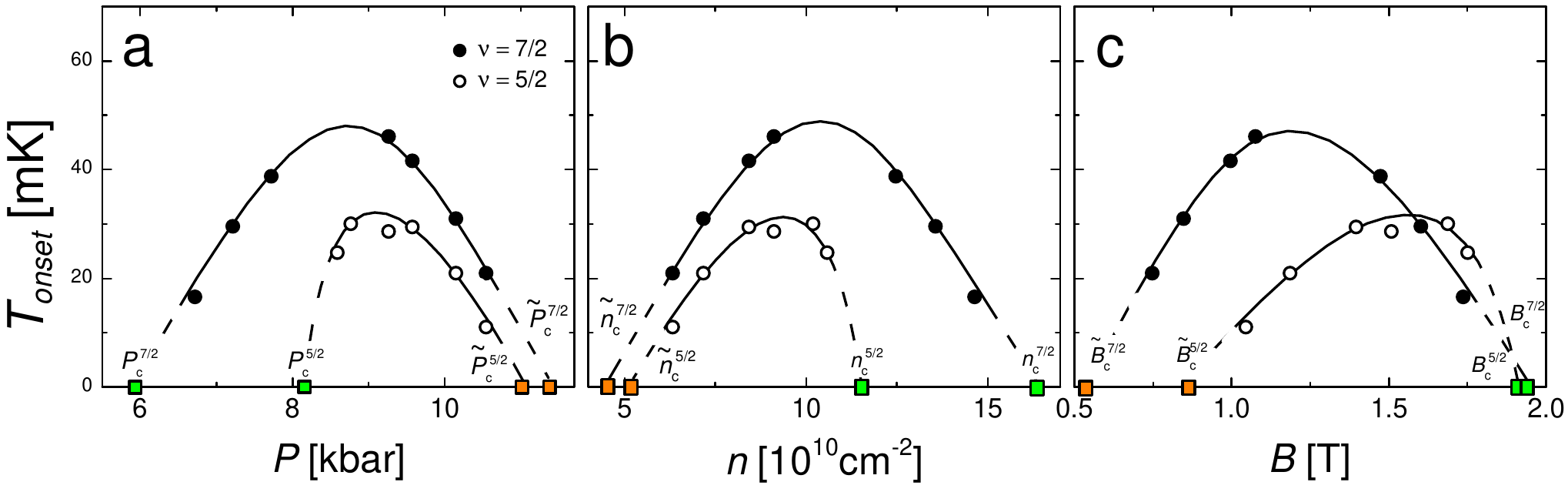}
\caption{
{\bf The dependence of the nematic onset temperature on various parameters.}
The onset temperatures of the nematic at $\nu = 5/2$ and $\nu = 7/2$ 
in Sample A is plotted as function of pressure (panel {\bf a}), electron density (panel { \bf b}), 
and magnetic field (panel {\bf c}). 
Green and orange squares are the estimated critical points of the 
paired-to-nematic and nematic-to-Fermi fluid quantum phase transitions, respectively.
Lines are guides to the eye.}
\label{fig-4}
\end{figure*}

{\bf Temperature dependence and the stability diagram.}
In order to understand the evolution of phases with pressure, we turn to
finite temperature measurements. We extract a characteristic energy scale of each ordered phase.
We define the onset temperature for the nematic $T_{onset}$ as the temperature at which $R_{xx}=2R_{yy}$ 
and the energy gap $\Delta$ of a FQHS by fitting the magnetoresistance to an activated expression
$ e^{-\Delta/2k_BT}$. The obtained values are summarized in Supplementary Tables 1 and 2.
By plotting these two quantities against pressure,
we obtain the stability diagrams in $P$-$T$ space shown in Fig.3. 
The stability diagram at $\nu=5/2$ has three regions \cite{Kate}. 
At low pressures, we observe a fractional quantum Hall ground state at $T=0$ and thermally
excited quasiparticles at finite $T$; the energy gap of the FQHS decreases with an increasing pressure.
At higher pressures we observe nematicity under a dome-like region.
At even higher pressures the nematic is destroyed into a featureless Fermi fluid. 
In our earlier work we argued that the simplest explanation for 
the sequence of the phases and of the stability diagram at $\nu=5/2$ is the existence of two
quantum phase transitions in the limit of $T=0$: one from a paired FQHS to the nematic occurring
at $P_c$, and another from the nematic to an isotropic Fermi fluid at $\tilde{P}_c$ \cite{Samkharadze,Kate}. 
Fig.3 reproduces this earlier result at $\nu=5/2$ in a sample of similar structure and of similar density, 
but cut from a different wafer \cite{Kate}.
Furthermore, the stability diagram at $\nu=7/2$, also shown in Fig.3, is qualitatively similar to that at $\nu=5/2$
as it also exhibits the same phases and the same two quantum critical points.

Our observation of competition of the FQHS and the nematic near the quantum critical point
highlights the importance of pairing in our experiments. Of the large number of
FQHSs forming in the second Landau level \cite{firstfivehalf,fivehalf,sevenhalf,Xia,Ethan}
only the paired FQHSs at $\nu=5/2$ and $7/2$ show the pressure induced transition to the
nematic. Indeed, the nematic in our pressurized samples does not develop at well-known filling factors,
such as the ones at $\nu=7/3$, $8/3$, $11/5$ or $14/5$, at which the ground state in the ambient
are FQHSs lacking pairing. Furthermore, in the parameter space accessed in our experiment,
we did not observe a paired-to-nematic quantum phase transition
at any other half-filled  Landau levels, such as at $\nu=9/2$ 
in the third Landau level or at $\nu=3/2$ in the lowest Landau level.
Taken together, these results establish the independence on the spin branch of the
stability diagram and of the paired-to-nematic 
quantum phase transition in the second orbital Landau level.

In the following we focus on the critical point of the paired-to-nematic quantum phase transition.
We estimate the critical pressure of the paired-to-nematic transition to be
half way between the highest pressure for the FQHS and the lowest pressure for the nematic.
We obtain $P_c^{5/2}=8.2 \pm 0.5$~kbar and $P_c^{7/2}=5.9 \pm 0.6$~kbar; 
these critical points are marked in Fig.3 by green squares. 
The critical pressure at $\nu=5/2$ is consistent with $7.8$~kbar, the value found in a similar sample 
\cite{Samkharadze,Kate}.
We attribute the difference of the two pressures to the 3\% difference in the density of the two samples and 
to variations due to room temperature cycling of the sample described in Methods. Strikingly,
the critical pressure $P_c^{7/2}=5.9$~kbar at $\nu=7/2$ is much
reduced from its value at $\nu=5/2$. We notice that in our sample the ratio of the critical 
pressures $P_c^{5/2}/ P_c^{7/2}=8.2/5.9 \approx 1.4$ is equal to the ratio of the two filling factors $7/5=1.4$.
This result suggests that pressure is not a primary driving parameter of the transition, but there may
be other ways to induce the same quantum phase transition. 
This hypothesis is not unreasonable since pressure tunes all band parameters, some of which
are discussed in Supplementary Note 1.
The quantity changing the most dramatically with pressure
is the electron density: it decreases linearly with pressure, reaching at 10~kbar
nearly 20\% of its value in the ambient \cite{p1,Samkharadze,Kate}. 
In Fig.4  we explore the premise of other driving parameters by plotting the nematic onset temperature against 
pressure, electron density, and magnetic field. Fig.4c is particularly significant, showing that in
Sample A the paired-to-nematic critical point at the two different filling factors
is at nearly the same magnetic field: $B_c^{5/2}=1.91$~T and $B_c^{7/2}=1.94$~T.

\begin{figure}[t!]  
\centering
\includegraphics[width=0.9\columnwidth]{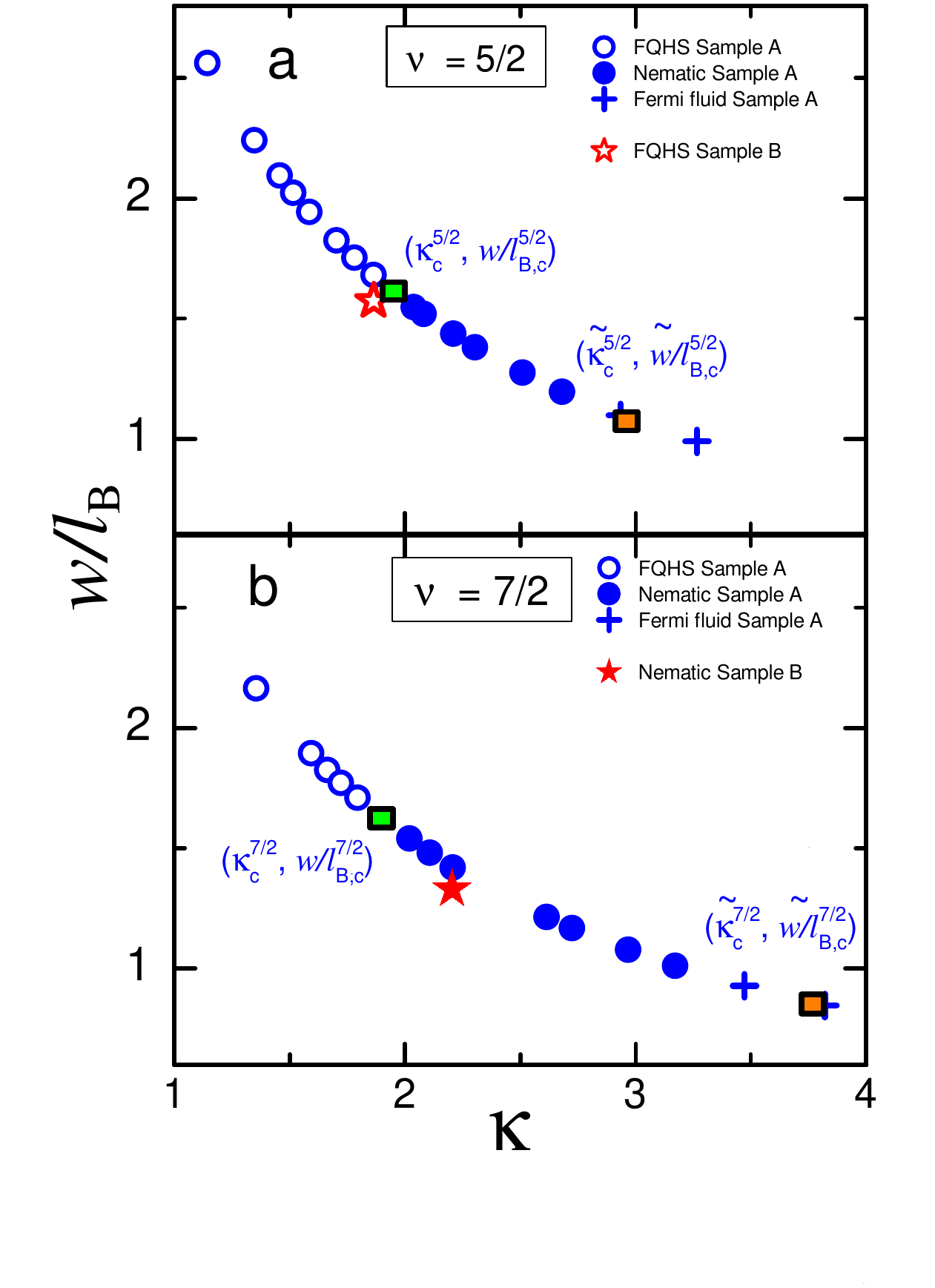}
\caption{
{\bf Sampling of the $\kappa$-$w/l_B$ space at half-filling.}
Parameters are calculated at $\nu=5/2$ (panel {\bf a}) and $\nu=7/2$ (panel {\bf b}).
Open symbols represent fractional quantum Hall states, closed ones nematics, and crosses
Fermi fluid ground states. Squares are quantum critical points for our pressurized Sample A.
Stars show parameters for Sample B, measured at ambient pressure.
}
\label{fig-5}
\end{figure}

The competition of the FQHS and of the nematic hinges on a delicate energy balance of these phases
near the quantum critical point. We propose that this phase competition is driven by the electron-electron
interaction, which in Sample A is tuned by pressure. The role of
the electron-electron interaction in stabilizing different ground states of the 
two-dimensional electron gas is well known \cite{pair4,wang}. 
In a realistic sample the electron-electron interaction is modified from its Coulomb expression
by the structure of the Landau levels \cite{llm0,llm1,llm2,llm2b,llm3,llm3b,llm4,llm5,llm6,Kennett}
and the finite thickness of the electron layer in the direction perpendicular to the plane of the electrons $w$
\cite{pair4,width1,width2,width3}.
These effects are encoded in two adimensional quantities: the Landau level mixing parameter
$\kappa=E_C/ \hbar \omega$ and the adimensional width of the electron layer $w/l_B$.
Here $E_C = e^2/( 4 \pi \epsilon l_B)$ is the Coulomb energy,
$\hbar \omega$ is the cyclotron energy, and $l_B = \sqrt{\hbar /e B}$ the magnetic length. 
The Landau level mixing parameter scales as $\kappa \propto m/\epsilon \sqrt{B}$, where $m$ is the effective
mass of electrons. Thus, in a given orbital Landau level and at fixed $m$, $\epsilon$, and $w$,
both $\kappa$ and $w/l_B$ are functions of the magnetic field only.
Under such constraints, therefore, the electron-electron interaction depends only on the magnetic field.
We conclude that the observation of a paired-to-nematic quantum critical point at both $\nu=5/2$
and $\nu=7/2$ at the same critical magnetic field may indeed be due to the tuning of the electron-electron interaction.
We think that in Sample A this interaction is tuned by the pressure through changing the electron density.
As we tune the pressure, in the $\kappa$-$w/l_B$ space we sample the curves shown in Fig.5.
At the critical pressure of the paired-to-nematic transition we find
$\kappa_c^{5/2}=1.95$, $w/l_{B,c}^{5/2}=1.62$ and $\kappa_c^{7/2}=1.90$, $w/l_{B,c}^{7/2}=1.63$,
nearly independent of the filling factor.
Here we took into account the pressure dependence of the effective mass and dielectric constant \cite{p1}.  
It is tempting to think of Fig.5 as a phase diagram. However, phase boundaries in this figure are expected 
to be significantly affected by disorder and by the lowest temperature reached. Nonetheless,
Fig.5 may serve as a guide to place constraints on the ordered phases. 
An expanded version of this figure, that includes published data obtained in samples in the ambient,
is shown in Supplementary Figure 3.

\begin{figure}[t]
\centering
\includegraphics[width=0.95\columnwidth]{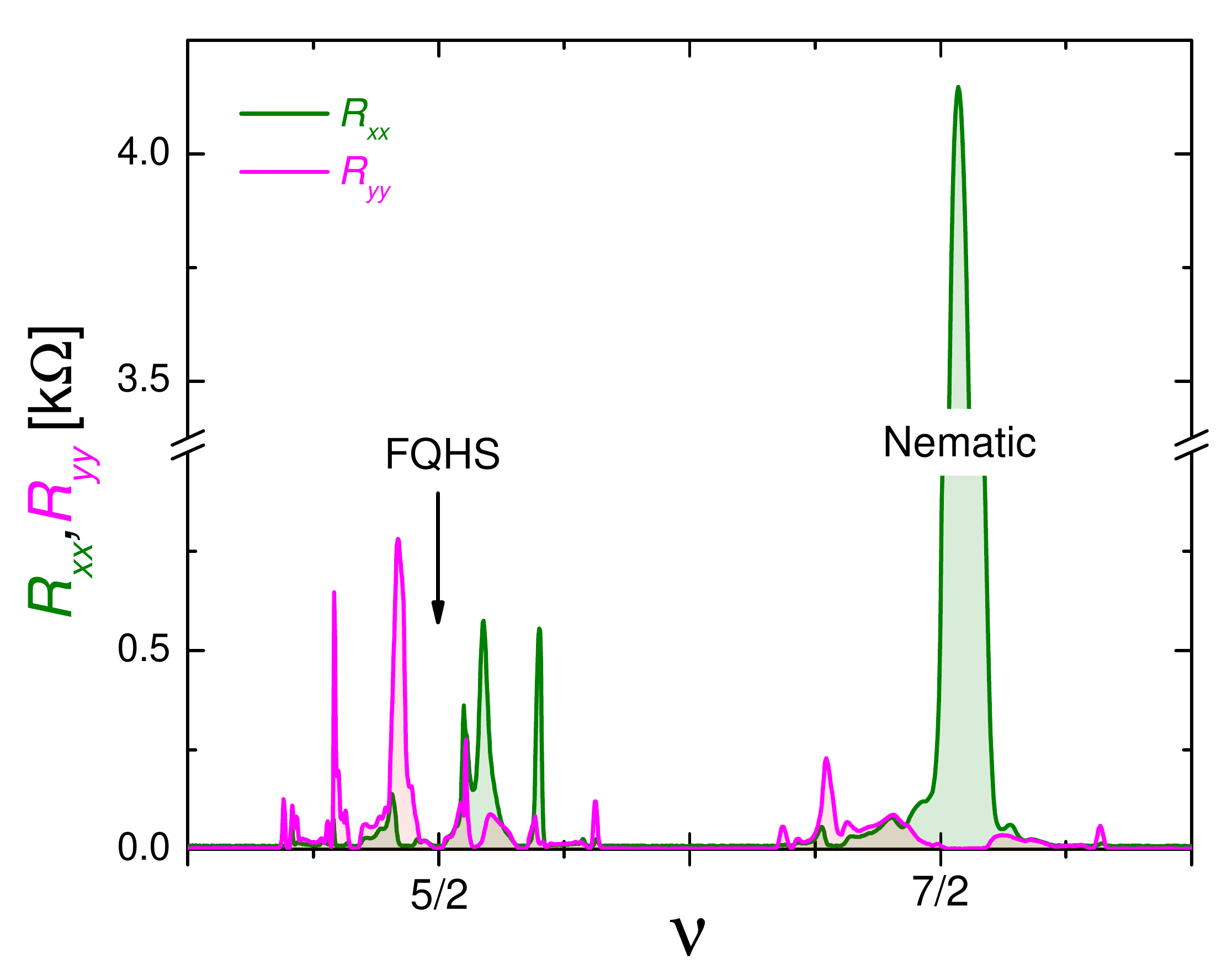}
\caption{ 
{\bf Magnetoresistance in the second Landau level of Sample B in the ambient.}
Traces are measured at ambient pressure and at $T \approx 4.5$ mK. 
The strong resistance anisotropy at $\nu=7/2$ shows a nematic ground state, while
at $\nu = 5/2$ we observe a FQHS.  The nature of the ground state
at these two filling factors is marked by text.
}
\label{fig-6}
\end{figure}

{\bf Measurements of a sample in the ambient.}
To test the relevance of electron-electron interactions, we investigate Sample B to be
measured at ambient pressure, but in which the electron-electron interaction was tuned near its value at
the quantum critical point. Sample B has the same width of the quantum well as Sample A, but 
it has a reduced density. By design, the density was picked in such a way that 
the parameters $\kappa$ and $w/l_B$ calculated at $\nu=7/2$ fall in the range of the nematic 
(shown as a red star in Fig.5). We note that data points for Sample B in Fig.5 are
slightly off the curve for Sample A since pressure corrections of the mass and dielectric strength are no longer needed.
Magnetoresistance traces for this sample, as measured with the sample mounted in a 
$^3$He immersion cell  \cite{Imm-cell}, are shown in Fig.6. At $\nu=7/2$ we indeed observe an extremely large
resistance anisotropy. Furthermore, at $\nu=5/2$ we observe a weak FQHS, consistent with
the $\kappa$ and $w/l_B$ parameters being just outside the range for the nematic. 
Hall resistance at $\nu=5/2$, shown in Supplementary Figure 4, is consistent with a FQHS.
Taken together, there is compelling evidence that the nematic phase is stabilized in the second orbital
Landau level at ambient pressure, when the electron-electron interaction is tuned via the parameters
$\kappa$ and $w/l_B$, to the stability range of the nematic. We emphasize that, according to our findings, the numerical
values of the critical parameters of the paired-to-nematic transition are valid only for $\nu=5/2$
and $7/2$ in the second orbital Landau level and are dependent on parameters such as
the width of the quantum well.

{\bf Discussion.}

It is interesting to note that in Sample A the nematic develops at $\nu=5/2$ for
pressures for which the electron density is in the range of $10.6-6.3 \times 10^{10}$~cm$^{-2}$.
Such densities have already
been accessed, but the nematic at $\nu=5/2$ was not observed \cite{ld1,ld2,llm3,Pan72}. 
Since samples from Refs.\cite{ld2,Pan72} had a wider quantum well than our samples,
the nematic in them either does not develop or it
forms at a yet unknown critical $\kappa$ and $w/l_B$ parameters.
The other two samples, however, had quantum wells of the same width as our samples \cite{ld1,llm3}.
In one of these samples densities necessary for the nematic, lower than
$10.6 \times 10^{10}$~cm$^{-2}$, have not been studied \cite{ld1}.
In the other $30$~nm quantum well sample the FQHS at $\nu=5/2$ is seen down to 
a density $12.5 \times 10^{10}$~cm$^{-2}$, but the nematic at $\nu=5/2$
was not seen at $9.5 \times 10^{10}$~cm$^{-2}$ \cite{llm3}. 
Possible reasons for the absence of the nematic in Ref.\cite{llm3} are 
disorder effects or effects due to the asymmetric shape of the wavefunction in the 
direction perpendicular to the plane of the electrons in gated samples. Resistance
anisotropy at $\nu=7/2$ was, however, observed in 60~nm quantum well sample having a 
density of $5 \times 10^{10}$cm$^{-2}$, providing an important clue on the
influence of the width of the quantum well \cite{Pan72}.
No data is available at $\nu=7/2$ in Refs.\cite{ld1,llm3}.

Interest in paired FQHSs has been recently rekindled by the discovery of FQHSs at even denominators
in electron gases confined to ZnO \cite{zno} and bilayer graphene hosts \cite{gr1,gr2}. 
However, in contrast to the GaAs system, in these hosts there is no evidence of the nematic.
The reason for the absence of the nematic in ZnO and bilayer graphene
is not currently known; disorder effects, a different crystal symmetry and/or a different electron-electron 
interaction may be at play. However, we cannot rule out
future observations of the nematic in these hosts. 
We will next compare the electron-electron interaction in these systems as parametrized by $\kappa$ and $w/l_B$.
The dielectric environment of the bilayer graphene encapsulated in boron nitride
is not well characterized; we will use $\epsilon \simeq 3.5$
and $m=0.05m_0$. For this host the parameters for the densities accessed fall in the
$\kappa \simeq 1.5 - 2.8$ and $w/l_B \simeq 0.03 - 0.06$ range \cite{gr2}. 
For the strongest $\nu=7/2$ FQHS developing in ZnO \cite{zno}, we find $\kappa \simeq 15$ and $w/l_B \simeq 0.3$. 
It is interesting to note that, in comparison to the GaAs system \cite{nod}, 
the even denominator FQHSs in bilayer graphene
develop at similar values of $\kappa$, but at much reduced value of $w/l_B$.
This is in sharp contrast with ZnO, in which the even denominator states
develop at an extremely large values of $\kappa$. We think that this opens the possibility 
that the nature of the even denominator FQHSs in ZnO may be fundamentally
different from those developing in GaAs or bilayer graphene.

Enhanced quantum fluctuations may have observable consequences close to the critical point.
A recent theory has examined the influence of the nematic fluctuations on the paired FQHS \cite{kim17}.
Nematic fluctuations may also influence the nematic phase itself in a description beyond the
mean field \cite{fogler,moessner}.
Our data show several anomalies close to the quantum critical point which may
be related to fluctuation effects.
One anomaly, shown in Fig.2c, is the resistance anisotropy at $\nu=7/2$
exceeds that at $5/2$. At fixed density and fixed temperature, a larger anisotropy typically develops
in the lower spin branch. For example, in the third orbital Landau level the anisotropy
observed at $\nu=9/2$ is larger than that at $\nu=11/2$ \cite{lilly,du}.
Other anomalies develop in Sample B, shown in Fig.6.
The resistance near $\nu=5/2$ is not isotropic in the vicinity of $\nu=5/2$ and data at $\nu \approx 2.42$
suggests a nematic which is not centered at half-filling. Furthermore, resistance anisotropy in the
upper spin branch is not exactly centered to $\nu=7/2$. Since the mean field approach
predicts a nematic centered at half-filling \cite{fogler,moessner}, we think that 
this approach is insufficient to describe the anomalies we see and that fluctuations are most likely at play.
Fluctuation effects stemming from the proximity
to the paired-to-nematic quantum critical point warrant further investigations.

In Fig.3 there is a second quantum phase transition at high pressures, from the nematic to an isotropic Fermi fluid. 
The critical pressures of this transition,
$\tilde{P}_c^{5/2}=11.0 $~kbar and $\tilde{P}_c^{7/2}=11.4$~kbar,
are  estimated by linear extrapolation to $T=0$ of the 
nematic onset temperatures forming at the two highest pressures.
These critical points are marked in Fig.3 by orange squares.
When comparing the critical values of different parameters at $\nu=5/2$ and $7/2$
which may drive the nematic-to-Fermi fluid transition
we find that, in contrast to the paired-to-nematic transition, this transition occurs at nearly the
same pressure, at values of the electron density close to each other
$\tilde{n}_c^{5/2} = 5.2 \times 10^{10}$~cm$^{-2}$  and  
$\tilde{n}_c^{7/2} = 4.5 \times 10^{10}$~cm$^{-2}$, but at very different magnetic fields.
The nematic onset temperature as function of these parameters is seen in Fig.4.
As discussed in Supplementary Note 2, at such low electron densities we expect that disorder 
effects do not permit nematic order. 
We thus think that the destruction of the nematic both at $\nu=5/2$ and $\nu=7/2$ at similar
electron densities is an indication that disorder became dominant. This idea is further supported by
data in Supplementary Figure 5, which depicts the suppression of the nematic at high pressures
in both the second and third Landau levels.

In summary, the observation of the pressure-driven quantum phase transition from a paired FQHS to the nematic
at both $\nu=5/2$ and $\nu=7/2$ Landau level filling factors
assures the independence of the spin branch of this transition in the second orbital Landau level.
Furthermore, by observing phases consistent with those at high pressure in a sample in the ambient, we have
shown that pressure is not the only driving parameter of this transition. 
Our observations suggest that tuning the electron-electron interactions,
as parametrized by Landau level mixing and adimensional width of the quantum well, play a critical role in driving the 
paired-to nematic phase transition. These results invite further investigations of the effects of fluctuations
developing near the quantum critical point on both the paired FQHS and on the nematic.

{\bf Methods.} 

{\bf Cryogenic measurements.}
Measurements were performed in a dilution refrigerator, using a standard low frequency lockin technique.
Magnetic fields up to 10~T were applied perpendicularly to the plane of the electron gas. 
Before cooling to low temperatures, samples were illuminated at 10~K using a red light emitting diode.

{\bf Details of the pressure cell and sample illumination.}
Sample A was cut to a $2 \times 2$~mm$^2$ size and was mounted in a pressure cell \cite{pcell}. 
The pressure transmitting fluid was an equal mixture of pentane and isopentane. In order to change pressure, 
the sample was warmed up to room temperature. After each room temperature cycling, the same
illumination technique was used. We estimate the lowest electronic temperature reached in this pressure cell is
about 12~mK.

{\bf Details of measurements under ambient pressure.}
Sample B was cut to a $4 \times 4$~mm$^2$ size and was measured in a $^3$He immersion cell \cite{Imm-cell}.
Using this cell we can thermalize electrons to temperatures below 5~mK. Details of the immersion
cell setup are found in Supplementary Note 3.

{\bf Acknowledgements.} Research at Purdue was supported by the US Department of Energy, 
Office of Basic Energy Sciences, Division of Materials Sciences and Engineering under the awards DE-SC0006671
(G.A.C. and M.J.M.) and DE-SC0010544 (Y.L.G.). K.A.S. acknowledges the 
Purdue Cagiantas Graduate Research Fellowship in Science.
L.N.P. and K.W.W. of Princeton University acknowledge the Gordon and Betty Moore Foundation
Grant No. GBMF 4420, and the National Science Foundation MRSEC Grant No. DMR-1420541.

\end{document}